\documentstyle[12pt]{article}

\input{epsf}

\newcommand{\sect}[1]{\setcounter{equation}{0}\section{#1}}

\def\RN{Reissner-Nordstr\"om}

\def\be{\begin{equation}}
\def\ee{\end{equation}}
\def\ba{\begin{eqnarray}\samepage}
\def\ea{\end{eqnarray}}

\font\twelvemsa=msam10 scaled 1200

\font\sevenmsa=msam7
\font\fivemsa=msam5
\newfam\msafam
\textfont\msafam=\twelvemsa
\scriptfont\msafam=\sevenmsa
\scriptscriptfont\msafam=\fivemsa
\def\msa{\ifcase\msafam 0\or1\or2\or3\or4\or5\or6\or7\or8\or9\or A\or B\or
C\or D\or E\or F\fi}

\font\twelvemsb=msbm10 scaled 1200

\font\sevenmsb=msbm7
\font\fivemsb=msbm5
\newfam\msbfam
\textfont\msbfam=\twelvemsb
\scriptfont\msbfam=\sevenmsb
\scriptscriptfont\msbfam=\fivemsb
\def\msb{\ifcase\msbfam 0\or1\or2\or3\or4\or5\or6\or7\or8\or9\or A\or B\or
C\or D\or E\or F\fi}

\font\twelveeuf=eufm10 scaled 1200

\font\seveneuf=eufm7
\font\fiveeuf=eufm5
\newfam\euffam
\textfont\euffam=\twelveeuf
\scriptfont\euffam=\seveneuf
\scriptscriptfont\euffam=\fiveeuf
\def\euf{\ifcase\euffam 0\or1\or2\or3\or4\or5\or6\or7\or8\or9\or A\or B\or
C\or D\or E\or F\fi}

\def\Bbb#1{\fam\msbfam#1}
\def\fraction#1#2{{\textstyle\frac{#1}{#2}}}
\def\half{\textstyle\frac{1}{2}}

\mathchardef\gapprox"3\msa26
\mathchardef\lapprox"3\msa2E

\begin{document}

\title{DYSON-PAIRS AND ZERO-MASS BLACK HOLES}

\author{G W Gibbons\\ \&\\D A Rasheed\thanks{Supported by EPSRC grant
no.~9400616X.}\\ \\DAMTP\\Silver Street\\Cambridge\\CB3 9EW}

\maketitle

\begin{abstract}
It has been argued by Dyson in the context of QED in flat spacetime
that perturbative expansions in powers of the electric charge $e$
cannot be convergent because if $e$ is purely imaginary then the
vacuum should be unstable to the production of charged pairs. We
investigate the spontaneous production of such Dyson pairs in
electrodynamics coupled to gravity. They are found to consist of pairs
of zero-rest mass black holes with regular horizons. The properties of
these zero rest mass black holes are discussed. We also consider ways
in which a dilaton may be included and the relevance of this to recent
ideas in string theory. We discuss accelerating solutions and find
that, in certain circumstances, the `no strut' condition may be
satisfied giving a regular solution describing a pair of zero rest
mass black holes accelerating away from one another. We also study
wormhole and tachyonic solutions and how they affect the stability of
the vacuum.
\end{abstract}

\renewcommand{\thepage}{ }
\pagebreak

\renewcommand{\thepage}{\arabic{page}}
\setcounter{page}{1}

\sect{Introduction}

Recently there has been much interest in the role of Bogomol'nyi
saturated states corresponding to extreme black holes in string
theory. One particularly interesting suggestion is that at special
points of the moduli space of string vacua these states may become
massless and the theory might thereby exhibit enhanced symmetry
\cite{HulTow}. Another related suggestion is that these states might
appear at points in the space of Calabi-Yau vacua associated to
singular geometries with conifold points \cite{Str}. Away from these
special points there is a region of moduli space where these
Bogomol'nyi black hole states are well described semi-classically by
classical solutions of the Einstein equations coupled to an
appropriate matter system, but presumably as one approaches a special
point the semi-classical description breaks down.

Nevertheless attempts have been made to find classical solutions
describing black holes with zero ADM mass
\cite{KalLin}-\cite{Ort}. These solutions are supersymmetric,
i.e. they admit Killing spinors, but they are nakedly singular. By the
Positive Energy Theorem for black holes these singularities are
inevitable because the matter sources satisfy the Dominant Energy
Condition.  This is a pity because there are many questions one would
like to ask about the properties of these putative zero-rest mass
black holes.  How do they move for instance?  The solutions given in
\cite{KalLin}-\cite{Ort} are static and at rest. This is not possible
for ordinary zero-rest mass particles.  What happens if these zero
rest mass black holes receive a small kick?  Are they stable? What
happens if one tries to accelerate them using an electric field? It is
difficult to answer questions like this using singular solutions
because the answers one obtains are dependent upon the boundary
conditions one imposes at the singularities.

The aim of this paper is to study a related but hopefully simpler
situation.  Consider, to begin with, ordinary electrodynamics.
Ignoring a possible theta term, the theory is specified by the Maxwell
Lagrangian
\be
 - { 1 \over 4 e^2} F_{\mu \nu} F^{\mu \nu}
\ee
where the electric charge $e$ may be said to label the possible vacua.
This is clear in string theory where it is an expectation value
\be
e^2 = \langle \exp 2 \phi \rangle
\ee
where $\phi$ is the dilaton. Studying the dependence of theory as a
function of the coupling constant $e$ is therefore the same as
studying the behaviour of the theory as one moves around the moduli
space of vacua. One immediate question is whether various properties
of the theory depend analytically upon the coupling constant. This
would be true in a neighbourhood of the origin if perturbation theory
in powers of $e^2$ were uniformly convergent. It was argued long ago
by Dyson \cite{Dys} that this cannot be so. The argument is by
contradiction. If the series converged in this way it would do so
everywhere within a small enough disc about the origin. In particular
perturbation theory would converge if $ e^2$ were negative. But in
such a world, like charges would attract and they would destablize the
vacuum.

In a gravity theory charged black holes, particularly the extreme
black holes, behave very much like charged particles and so this
beautiful argument of Dyson suggests looking at what happens to black
holes in a world in which, instead of the standard
Lagrangian\footnote{In this paper we use units such that
$G=c=4\pi\varepsilon_0=1$}
\be
{1\over 16\pi} \left(R - F_{\mu\nu}F^{\mu\nu}\right),
\ee
we change the sign of the electromagnetic contribution and use
\be
{1\over 16\pi} \left(R + F_{\mu\nu}F^{\mu\nu}\right)
\ee   
instead. More generally we could consider a theory with a dynamical
scalar $\sigma$.  The action would be
\be
{1\over 16\pi} \left(R - 2(\partial\sigma)^2 +
e^{-2a\sigma}F_{\mu\nu}F^{\mu\nu}\right),
\ee
where $a$ is a dimensionless coupling constant. Note that the scalar
$\sigma$ has positive kinetic energy. If $a=0$ we may consistently set
$\sigma=0$ and we get back to the usual Einstein-anti-Maxwell case.
If $a= \sqrt 3$ we get anti-Kaluza-Klein theory in which the extra
dimension is timelike rather than the usual spacelike case. Our
results may therefore also be relevant to theories with more than one
time direction \cite{Ben}. The case $a=1$ corresponds to string
theory. Formally the action above is obtained by setting
\be
\sigma = \phi + i {\pi \over 2a} 
\ee
in the usual action. Because this is just a displacement of $\sigma$
it leaves the kinetic term unchanged.  It may be of interest to note
that Lindstrom and Rocek have drawn attention to vector fields with
negative kinetic energies in Yang-Mills theories with zero mass
monopoles \cite{LinRoc}.

Another application of these ideas is to the field outside a
fundamental string in four spacetime dimensions \cite{DabGib} (or more
generally an $(n-3)$-brane in $n$ spacetime dimensions). The dilaton
behaves as
\be
e^{-2\phi} = 1 - 8\mu\ln r
\ee
where $r$ is the radial distance from the core of the string.  Note
that the $\phi$ used here is twice that used in \cite{DabGib}. At
sufficiently large transverse distances $e^{2 \phi}$ becomes negative
which means that the effective string coupling constant becomes pure
imaginary.  The metric becomes singular at $r=e^{1 \over 8G \mu}$ and
so it cannot be interpreted in a straightforward way as an ordinary
horizon. In the Kaluza-Klein context a similar phenomenon is
encountered and one may have an ordinary horizon in the higher
dimensional space because the Killing field one is reducing on
switches from being spacelike to being timelike.

Of course one may, if one wishes, consider a scalar field with
negative kinetic energy. This also produces some exotic solutions even
without an electromagnetic field, including an Einstein-Rosen bridge
with $g_{00}=-1$ and thus no horizon separating the two sides. This is
discussed briefly in section~2.2 and in more detail in section~5.

In what follows we shall sometimes use, as we did above, the prefix
``anti" to describe quantities associated to an electromagnetic field
of the opposite sign from usual.  Thus, in the next section, we will
consider black holes with ``anti-charge". Of course, like anti-charges
attract rather than repel, as they do ordinarily. If we take the
conventional sign for the scalar field kinetic term, then it also
leads to an attractive force. As we shall see this means that extreme
black holes are not possible in these theories. Moreover none of the
solutions we shall discuss admit Killing spinors. Another difference
is that the solutions are non-singular outside a regular event
horizon. Thus the phenomena we are about to describe are not precisely
the same as those considered in \cite{HulTow}-\cite{Ort}. However the
solutions in this paper do share the common feature with those of
\cite{KalLin}-\cite{Ort} that the total 4-momentum vanishes. It seems
quite possible therefore that they may throw some light on the
dynamical behaviour of such objects.

\sect{Static Zero Mass Black Holes}

The relevant metrics and fields for electrically charged solutions in
Einstein-anti-Maxwell-dilaton theory may be obtained from the usual
case \cite{GarGar,GibMae} by setting the charge to be pure
imaginary. The four-metric is
$$
ds^2 = -\left(1-{r_+\over r}\right) \left(1-{r_-\over
r}\right)^{1-a^2\over 1+a^2}dt^2 + \left(1-{r_+\over r}\right)^{-1}
\left(1-{r_-\over r}\right)^{a^2-1\over 1+a^2}dr^2
$$
\be
+r^2 \left(1-{r_-\over r}\right)^{2a^2\over 1+a^2}(d\theta^2 +
\sin^2\theta d \phi^2),
\ee
with scalar field
\be
e^{\sigma} = \left(1-{r_-\over r}\right)^{a\over 1+a^2} 
\ee
and Maxwell field
\be
F = {Q\over r^2} dt\wedge dr.
\ee
For the {\em magnetic} case the metric is the same but the sign of the
scalar field $\sigma$ must be reversed and the Maxwell field becomes
$F=P\sin\theta d\theta\wedge d\phi$. The ADM mass $M$ is
\be
M= {1 \over 2} \left(r_+ + {1-a^2\over 1+a^2}r_-\right)
\ee
and the anticharge $Q$ is given by
\be
|Q| = \sqrt{-r_+r_-\over 1+a^2}.
\ee
The scalar charge $\Sigma$ is given by
\be
\Sigma = - {ar_-\over 1+a^2},
\ee
thus
\be
M^2 + \Sigma ^2 + Q^2 = {1\over 4} (r_+-r_-)^2.
\ee

From the spacetime point of view these solutions will have a regular
event horizon at $r=r_+$ and behave like black holes provided that
$r_+$ is positive and $r_-$ is {\em negative} (the usual case has both
positive but with $r_- \le r_+$). In the case $a=0$, $r_-$ would be a
Cauchy horizon, or more generally it is a singularity, but because
$r_-$ is negative one passes through the horizon $r_+$ to reach the
singularity at $r=0$ without ever reaching $r_-$. The causal structure
is thus the same as the Schwarzschild solution. For the same reason
there are no extreme holes.  One may also check that one cannot
analytically continue the usual multi-solutions to have imaginary
values of the charge, and so anti-gravity is excluded.\footnote{In
fact it may be possible to have anti-gravity in a limited sense~: in
principle, 2 such black holes with $Q=M$ and electric charges of {\em
opposite} signs should be able to remain in a static equilibrium since
unlike charges repel in this theory. It is obvious, however, that such
a static equilibrium is not possible for more than 2 black holes,
since they cannot all have charges of different signs from one
another, and so this is not true anti-gravity.}

The behaviour of the solutions depends in an essential way on whether
$a^2$ is less or greater than unity.  In the former case we may, by
taking $r_- $ sufficiently negative, violate the positive mass
theorem. In fact, by choosing
\be
r_+ = {a^2-1\over 1+a^2}r_-,
\ee
we can obtain a solution with vanishing total ADM mass.  One may ask
what happens if $r$ is allowed to be negative.  One then has an
asymptotically flat spacetime with negative ADM mass containing a
naked singularity at $r=r_-$.

By contrast if $a^2\ge 1$ and $r$ is positive then the ADM mass $M$ is
always positive.  An explanation for this fact is presumably that the
larger the coupling constant $a$, the greater is the positive
contribution of the scalar field to the total energy, and if $a^2\ge
1$ this overwhelms the negative contribution of the negative energy
vector field. In particular, in the case of anti-Kaluza-Klein theory
($a^2=3$), the black holes always have positive mass.

The case $a=-\sqrt 3$ is associated with the five dimensional metric
of signature $+++--$:
$$
ds^5 = -\left(1-{r_-\over r}\right)^{\mp 1} \left(d\tau + 2A_\mu
dx^\mu\right)^2
$$
$$
 + \left(1-{r_-\over r}\right)^{\pm{1\over 2}} \left\{
-\left(1-{r_+\over r}\right) \left(1-{r_-\over r}\right)^{-{1\over
2}}dt^2 + \left(1-{r_+\over r}\right)^{-1}
\left(1-{r_-\over r}\right)^{1\over 2}dr^2 \right\}
$$
\be
+r^2 \left(1-{r_-\over r}\right)^{3\over 2} (d\theta^2 + \sin^2\theta
d\phi^2)
\ee
where the upper sign corresponds to the electric case and the lower to
the magnetic case and the vector field $A_\mu$ must be chosen
accordingly. In either case, since for $r>0$ the metric component
$g_{\tau \tau}$ never vanishes, the regularity properties are the same
as for the four dimensional metric.

In fact, for $a^2\ge 1$, one can show that $M$ is not just positive
but it is bounded below by a positive quantity proportional to the
charge~:
\be
M \ge |Q|\sqrt{a^2-1}.
\ee
This is reminiscent of the Bogomol'nyi mass bound in ordinary EMD
theory, however, the solutions saturating this bound are not extreme
in the usual sense. They have 2 distinct horizons $r_\pm$ satisfying
\be
r_+ = {1-a^2\over 1+a^2}r_-.
\ee
The fact that these solutions are not extreme, i.e.\ $r_+\ne r_-$,
will have profound implications when we come to study their
thermodynamic properties. It will be seen that they have non-zero
Hawking temperature and the surface area of the event horizon is
non-zero. Hence one might expect that they should lose mass by
radiating neutral particles resulting in a decrease in the ratio
$M\over|Q|$. However, no static solutions exist with
${M\over|Q|}<\sqrt{a^2-1}$~! This will be discussed in more detail in
Section~4.

The main conclusion is that if one reverses the sign of the coupling
of the electromagnetic kinetic term in the action then static zero
rest mass black hole solutions which are non-singular outside a
regular event horizon are indeed possible for $a^2<1$. Since they are
spherically symmetric the total spatial momentum vanishes as well as
the mass and so they have vanishing total 4-momentum. In this respect
they should be distinguished from conventional massless particles
which have a non-zero but lightlike total 4-momentum. They should also
not be confused with tachyonic excitations which are associated with a
spacelike 4-momentum.

It is also interesting to consider what happens if the kinetic energy
of the scalar field is taken to be negative. In this case, the scalar
field produces a repulsive force and so extreme solutions may be
possible and also multi-centre solutions satisfying a force balance,
depending on the strength of the dilaton coupling $a$ and the sign of
the Maxwell kinetic term. If the Maxwell kinetic term is given the
opposite sign, then zero rest mass black holes again become a
possibility.
 
It has become conventional to use the abbreviation EMD for the
standard Einstein-Maxwell-dilaton theory in which both the Maxwell
field and the scalar are given positive kinetic energies. For our
purposes, it is therefore convenient to introduce the obvious
abbreviations E$\overline{\rm M}$D, EM$\overline{\rm D}$ and
E$\overline{\rm MD}$ for Einstein-anti-Maxwell-dilaton,
Einstein-Maxwell-anti-dilaton and Einstein-anti-Maxwell-anti-dilaton
theories respectively. We will now derive and investigate a class of
static spherically symmetric black hole solutions with electric charge
in each of these theories. These black holes will form a 2 parameter
family of solutions labelled by their mass $M$ and electric charge
$Q$, with the scalar dilaton charge $\Sigma$ being determined by $M$
and $Q$. We will also find a new wormhole solution which lies outside
the class of black hole solutions.

\subsection{EM$\overline{\bf D}$ Theory}

Here the Maxwell field is given the usual sign in the action but the
scalar is given negative kinetic energy~:
\be
{1\over 16\pi} \left(R + 2(\partial\sigma)^2 -
e^{-2a\sigma}F_{\mu\nu}F^{\mu\nu}\right).
\ee
Static spherically symmetric black holes solutions may be obtained by
making use of internal symmetries of the dimensionally reduced
3-dimensional action. Writing the metric in the form
\be
ds^2 = -e^{2u}dt^2 + e^{-2u}h_{ij}dx^idx^j,
\ee
where $i,j$ run from 1 to 3, and taking the Maxwell field strength
2-form to be the exterior derivative of the 1-form vector potential
$A=\phi dt$ where $\phi(x^i)$ is the electric potential, the system
may be described by the 3-dimensional non-linear $\sigma$-model action
\be
\int d^3x\sqrt{h} \left\{R(h) - 2(\partial u)^2 + 2(\partial\sigma)^2
+ 2e^{-2u-2a\sigma}(\partial\phi)^2\right\}.
\label{3DAct}
\ee
Charged solutions may thus be generated from neutral ones using the
symmetries of this action. The continuous symmetries of the action are
given infinitesimally by the following 4 Killing vectors acting on the
internal target space~:
$$
k^{(1)} = {\partial\over\partial\phi} ,\qquad k^{(2)} =
\phi{\partial\over\partial\phi} + {1\over
a}{\partial\over\partial\sigma} ,\qquad k^{(3)} =
{\partial\over\partial u} + \phi{\partial\over\partial\phi},
$$
\be
k^{(4)} = 2\phi{\partial\over\partial u} +
\left[e^{2u+2a\sigma}+(1-a^2)\phi^2\right]{\partial\over\partial\phi}
- 2a\phi{\partial\over\partial\sigma}.
\ee
$k^{(4)}$ is the generator of the required anti-dilaton-Harrison
transformation which produces electrically charged solutions from
neutral ones.

Alternatively, a class of static black holes may be obtained from the
usual dilaton black holes by making the substitutions
$\sigma\rightarrow i\sigma$ and $a\rightarrow -ia$. The new scalar
field $\sigma$ and $a$ are required to be real and so in the original
solution they must be made pure imaginary by analytic
continuation. The new metric is thus obtained simply by the
replacement $a^2\rightarrow -a^2$~:
$$
ds^2 = -\left(1-{r_+\over r}\right) \left(1-{r_-\over
r}\right)^{1+a^2\over 1-a^2}dt^2 + \left(1-{r_+\over r}\right)^{-1}
\left(1-{r_-\over r}\right)^{1+a^2\over a^2-1}dr^2
$$
\be
+r^2 \left(1-{r_-\over r}\right)^{2a^2\over a^2-1}(d\theta^2 +
\sin^2\theta d \phi^2).
\ee
The scalar field becomes
\be
e^\sigma = \left(1-{r_-\over r}\right)^{a\over a^2-1}
\ee
and the Maxwell field is
\be
F = {Q\over r^2}dt\wedge dr,
\ee
where
\be
Q = \sqrt{r_+r_-\over 1-a^2}.
\ee
Thus for $a^2<1$ we must have $0\le r_-\le r_+$ and for $a^2>1$,
$r_-\le 0\le r_+$. Clearly then extreme solutions are only possible if
$a^2<1$, and they satisfy
\be
M^2 = {Q^2\over 1-a^2}.
\ee
The scalar charge is given by
\be
\Sigma = {ar_-\over 1-a^2}
\ee
and the ADM mass is
\be
M = {1\over 2}\left(r_++{1+a^2\over 1-a^2}r_-\right).
\ee
Thus we have
\be
M^2 - \Sigma^2 - Q^2 = \fraction{1}{4}(r_+-r_-)^2
\ee
and so these black holes cannot have zero mass.

The solutions for $a^2=1$ require more careful consideration. As we
take the limit $a^2\rightarrow 1$, it is clear that various metric
components will blow up, unless $r_-\rightarrow 0$. However, we may
obtain finite results by setting $r_-=|\Sigma|(1-a^2)$ and keeping
$\Sigma$ constant as we take the limit~:
$$
ds^2 = -\left(1-{r_+\over r}\right)e^{-2|\Sigma|/r}dt^2 + \left(1-{r_+\over r}\right)^{-1}e^{2|\Sigma|/r}dr^2
$$
\be
+ r^2e^{2|\Sigma|/r}(d\theta^2+\sin^2\theta d\phi^2).
\ee
The scalar field becomes
\be
\sigma = {\Sigma\over r},
\ee
where the sign of $\Sigma$ is the same as the sign of $a=\pm 1$ and
the Maxwell field is
\be
F = {\sqrt{|\Sigma|r_+}\over r^2}dt\wedge dr.
\ee
The ADM mass is now
\be
M = |\Sigma| + \half r_+
\ee
which is always positive.

For $a^2>1$ the causal structure is the same as for the Schwarzschild
solution, with a spacelike singularity at $r=0$ hidden behind a
regular event horizon at $r=r_+$. The singularity at $r=r_-$ is
unreachable, since $r_-\le 0$ in this case. For $a^2\le 1$, the
singularity at $r=r_-\ge 0$ is null (except in the Einstein-Maxwell
case, $a=0$, when it becomes the inner Cauchy horizon).

\subsection{E$\overline{\bf MD}$ Theory}

If we allow both the scalar and the Maxwell fields to have negative
kinetic terms, then we expect once more to find regular zero rest mass
black holes. The action is
\be
{1\over 16\pi} \left(R + 2(\partial\sigma)^2 +
e^{-2a\sigma}F_{\mu\nu}F^{\mu\nu}\right)
\ee
and the solutions may be obtained from those above by taking $Q$ to be
pure imaginary. The metric is the same but now
\be
Q = \sqrt{r_+r_-\over 1-a^2},
\ee
so if $a^2<1$, $r_-\le 0$ and if $a^2>1$, $r_-\ge 0$. Extreme solutions
satisfy
\be
M^2 = {Q^2\over a^2-1}
\ee
and so are only possible for $a^2>1$.

Zero mass black holes are now possible once more, if we set
\be
r_+ = {1+a^2\over a^2-1}r_-
\ee
for $a^2\ne 1$. The case $a^2=1$ must be checked separately. Setting
$r_-=|\Sigma|(a^2-1)$ and taking the limit $a^2\rightarrow 1$ we
obtain the metric
$$
ds^2 = -\left(1-{r_+\over r}\right)e^{2|\Sigma|/r}dt^2 + \left(1-{r_+\over r}\right)^{-1}e^{-2|\Sigma|/r}dr^2
$$
\be
+ r^2e^{-2|\Sigma|/r}(d\theta^2+\sin^2\theta d\phi^2),
\ee
with
\be
F = {\sqrt{|\Sigma|r_+}\over r^2}dt\wedge dr
\ee
and
\be
\sigma = {\Sigma\over r},
\ee
where the sign of $\Sigma$ is now opposite to that of $a$. The ADM
mass is therefore
\be
M = \half r_+ - |\Sigma|
\ee
which may become zero or negative.

Thus we see that zero rest mass black holes, non-singular outside a
regular event horizon, violating the Positive Energy Theorem, exist in
E$\overline{\rm M}$D theory for $a^2<1$ and in E$\overline{\rm MD}$
theory for all $a$, but not in EM$\overline{\rm D}$ theory. That is
not to say that the Positive Energy Theorem holds in EM$\overline{\rm
D}$ theory. Since the scalar field has negative kinetic energy, it
violates the Dominant Energy Condition (in fact it violates the Weak
Energy Condition) and so we expect counter-examples to the Positive
Energy Theorem. One such counter-example with $F_{\mu\nu}=0$ may be
obtained from the Schwarzschild solution by using the discrete duality
between the metric function $u$ and the dilaton $\sigma$ in the action
(\ref{3DAct}) with $\phi=0$. The resulting solution describes a {\em
transparent massless} wormhole~:
$$
ds^2 = -dt^2 + dr^2 + (r^2+c^2)(d\theta^2+\sin^2\theta d\phi^2),
$$
\be
\sigma = \tan^{-1}\left({r\over c}\right).
\label{Worm}
\ee
Since $F_{\mu\nu}=0$, this is a solution of {\em both} EM$\overline{\rm
D}$ and E$\overline{\rm MD}$ theory, violating the Positive Energy
Theorem. This is discussed in more detail in section~5.

It is interesting to ask whether such solutions are stable. Presumably
they are not. In the case of zero mass black holes, some insight into
this question may be gained by studying their thermodynamic properties
and also by considering accelerating solutions.

\sect{Accelerating Solutions}

In order to see how these black holes move we turn to the
dilaton-C-metrics \cite{KinWal,DowGau}. In E$\overline{\rm M}$D theory these have the
form
\be
ds^2 = {1\over A^2(x-y)^2} \left\{
F(x) \left[G(y)dt^2 - {dy^2\over G(y)}\right] +
F(y) \left[{dx^2\over G(x)} + G(x)d\phi^2\right]
\right\}.
\ee
In the {\em magnetic} case the scalar field is 
\be
e^{2a \sigma} = {F(x)\over F(y)}
\ee
where
\be
F(u) \equiv (1+r_-Au)^{2a^2\over 1+a^2}
\ee
and 
\be
G(u) \equiv \left[1-u^2(1+r_+A u)\right] (1+r_-Au)^{1-a^2\over 1+a^2}.
\ee
The vector field is given by
\be
A = \sqrt{-r_+r_-\over 1+a^2}xd\phi.
\ee
The electrically charged solutions may be obtained by a generalized
duality transformation. This transformation leaves the metric
unchanged. As before we are interested in the case when $r_+$ is
positive but $r_-$ negative. We take the acceleration parameter $A$ to
be positive. The behaviour of the solutions depends upon the cubic
\be
{\overline G}(u) \equiv 1-u^2(1+r_+A u).
\ee
Because $r_-$ is negative the ordering of the roots is different from
the standard one and so our labelling is different. For
$0<r_+A<{2\over\sqrt{27}}$ the cubic ${\overline G}(u)$ has three real
roots, which we label in increasing order~:
\be
-{1\over Ar_+} < u_1 < -\sqrt3 < u_2 <-1 \quad\mbox{and}\quad {\sqrt
3\over 2} < u_3 < 1.
\ee
Thus roots $u_1$ and $u_2$ are negative and $u_3$ is positive. In
addition, if $a^2 <1$ then $G(u)$ has another zero corresponding to
the root of the function
\be
{\overline F}(u) = 1+r_-Au
\ee
i.e.\ $u=u_4=-{1\over Ar_-}$.

If we interpret $x$ as an angular coordinate then the requirement that
$g_{\phi \phi} >0$ and that it have finite range dictates it must vary
between the least negative and the least positive root of $G(u)$, i.e $
x \in [u_2, u_3]$.  The end points of this interval will be regular
axes of symmetry free of conical singularities if and only if the `no
strut' condition
\be
G^\prime(u_2)+G^\prime(u_3)=0
\label{NoStrut}
\ee
holds. We are assuming of course that $(r_+,r_-,A)$ are chosen so that
\be
u_3<u_4 = -{1\over r_-A} .
\ee
Because $u_3 <1$ it suffices that $|r_-|<{1\over A}$ for this to be
true.

If the coordinate $t$ is to be timelike then the radial variable $y$
must range between the two negative or the two positive roots of
$G(u)$. In the limit $A\rightarrow 0$, the coordinate transformation
$y\rightarrow -{1\over Ar}, x\rightarrow\cos\theta, t\rightarrow At,
\phi\rightarrow\phi$ gives the E$\overline{\rm M}$D black hole
solutions in the usual coordinates. Thus $y=u_1$ corresponds to the
black hole event horizon and, for $A\ne 0$, $y=u_2$ is the
acceleration horizon. Therefore we choose $y\in[u_1,u_2]$.  The
resulting solution will be asymptotically flat.  Because it is
boost-invariant the total ADM 4-momentum vanishes.  The solution may
be thought of as a pair of uniformly accelerating black holes with
`mass parameter' $M$ and anti-charge $Q$. However it is better to
think of $M$ as a measure of the size of the horizon. As in the case
of the static black holes considered in the last section, the Cauchy
horizon/singularity at $r_-$ cannot be reached because $r_-<0$ and
there is a singularity at $r=0$. The causal structure is thus
equivalent to that of the usual vacuum C-metric \cite{KinWal}.

It is well known that if the vector field has the usual sign in the
Lagrangian then it is impossible to satisfy the regularity condition
(\ref{NoStrut}) for any value of the electric charge, including zero,
and thereby eliminate the conical singularities. Indeed if this were
possible then the solutions would violate the Positive Mass Theorem
for black holes. Moreover one might then expect to be able to pick the
acceleration parameter of the solution so as to render the surface
gravities of the acceleration horizon and the event horizon equal. If
this were possible it would be possible to analytically continue the
time parameter to give a regular Euclidean instanton solution
(periodic in imaginary time) which would mediate the instability of
Minkowski spacetime.  However, as we shall see, the situation is
rather different if the vector field enters the Lagrangian with the
opposite sign. This is not surprising because we have seen in the
previous section that static non-singular configurations with
vanishing total 4-momentum exist in such theories.

The simplest case to consider is the Einstein-anti-Maxwell one, $a=0$.    
Then the function $G(u)$ becomes (after a suitable linear coordinate
transformation of $x$ and $y$, see \cite{KinWal}) the quartic :
\be
G(u)= 1 -u^2 -2MA u^3 + Q^2 A^2 u^4. 
\ee
Clearly if we set $M=0$ this quartic polynomial will be symmetrical
about the origin and the no strut condition is automatically satisfied
by symmetry.  Thus we have a regular Lorentzian solution representing
two uniformly accelerating black holes with equal and opposite charges
in an asymptotically flat spacetime with vanishing total ADM
mass. Note that ultimately the black holes will approach the velocity
of light.

Because the solution without an applied electric field has no conical
singularity, adding an electric field should produce a solution with a
conical singularity. If this is correct it indicates that trying to
accelerate a zero rest mass black hole with an electric field cannot
give rise to a constant acceleration.

What about the Instanton? We need to check to see whether we can
choose $QA$ so that the surface gravities of the event horizon and the
acceleration horizon can be equal.  This requires that the slopes at
the two roots $u=u_1$ and $u=u_2$ be equal in magnitude and opposite
in sign. This can only happen in the limiting case that they coincide
$u_1=u_2$.  This requires that the charge and acceleration be related
by
\be
|Q|={ 1 \over 2} A.
\ee
Thus we see that although accelerating solutions without conical
singularities exist in this case, there are no instanton solutions
which would mediate the decay of the vacuum. We will discuss vacuum
decay in more detail in section~7.

The situation with a scalar field present is only slightly more
complicated. Remaining with E$\overline{\rm M}$D theory, so the scalar
has positive kinetic energy, we recall that zero rest mass black holes
were possible only if $a^2<1$. Thus we expect to be able to satisfy
the no strut condition only if $a^2<1$ and obtain a regular solution
representing a pair of zero rest mass E$\overline{\rm M}$D black holes
accelerating away from one another.

Writing the function $G(u)$ as a product of its factors, we have
\be
G(u) = C(u-u_1)(u-u_2)(u-u_3)(u-u_4)^p
\ee
where
\be
p = {1-a^2\over 1+a^2}
\ee
and $u_1, u_2, u_3, u_4$ are ordered~:
\be
u_1 < u_2 < u_3 < u_4.
\ee
The no strut condition then becomes
\be
(u_2-u_1)(u_4-u_2)^p = (u_3-u_1)(u_4-u_3)^p.
\ee
We may solve this to express $u_4$, say, in terms of the other roots~:
\be
u_4 = {u_3(u_3-u_1)^{1\over p}-u_2(u_2-u_1)^{1\over
p}\over(u_3-u_1)^{1\over p}-(u_2-u_1)^{1\over p}}.
\ee
Finally, for consistency, we need to check that $u_4>u_3$. If $p>0$
(i.e.\ $a^2<1$), then the denominator of the expression for $u_4$ is
positive and we can multiply up. The inequality $u_4>u_3$ then
simplifies to
\be
(u_3-u_2)(u_2-u_1)^{1\over p} > 0,
\ee
which holds automatically. Thus the no strut condition can be made to
hold if $a^2<1$. If $a^2>1$ (i.e.\ $p<0$), the denominator in the
expression for $u_4$ is negative, all inequalities must be reversed on
multiplying up and the condition $u_4>u_3$ cannot be satisfied. In the
special case $a^2=1$, $p=0$ and the no strut condition reduces to
$u_2=u_3$.

Thus, as predicted, the conical singularity in the E$\overline{\rm
M}$D C-metric may only be removed if $a^2<1$, giving a regular
solution representing a pair of zero rest mass black holes with equal
and opposite charges accelerating away from one another.

\subsection{EM$\overline{\bf D}$ Accelerating Solutions}

The EM$\overline{\rm D}$ C-metric may be obtained from the usual
dilaton C-metric by the replacement $\sigma\rightarrow i\sigma$,
$a\rightarrow -ia$. The form of the metric remains unchanged but the
functions $F$ and $G$ become
\ba
F(u) &=& (1+r_-Au)^{2a^2\over a^2-1}, \nonumber \\
G(u) &=& \left[1-u^2(1+r_+Au)\right](1+r_-Au)^{1+a^2\over 1-a^2}.
\ea
The scalar field is again given by
\be
e^{2a\sigma} = {F(x)\over F(y)}
\ee
and the vector potential 1-form is
\be
A = \sqrt{r_+r_-\over 1-a^2}xd\phi.
\ee
For $a^2<1$, $0\le r_-<r_+$ and so the new ordering of the roots of
$G(u)$ becomes
\be
u_4 = -{1\over Ar_-} < u_1 < u_2 < u_3.
\ee
As before we restrict $x$ to lie between $u_2$ and $u_3$ so that the
metric component $g_{\phi\phi}$ is positive and $y\in[u_1,u_2]$. The
no strut condition $G^\prime(u_2)+G^\prime(u_3)=0$ is equivalent to
\be
(u_2-u_1)(u_2-u_4)^p = (u_3-u_1)(u_3-u_4)^p
\ee
where
\be
p = {1+a^2\over 1-a^2} > 0.
\ee
Thus solving for $u_4$, we obtain
\be
u_4 = {u_3(u_3-u_1)^{1\over p}-u_2(u_2-u_1)^{1\over
p}\over(u_3-u_1)^{1\over p}-(u_2-u_1)^{1\over p}}.
\ee
For consistency, we require that $u_4<u_1$ which, after multiplying
up, is equivalent to
\be
(u_3-u_1)^{1+{1\over p}} < (u_2-u_1)^{1+{1\over p}}.
\ee
But, since $1+{1\over p}={2\over 1+a^2}>0$, this inequality does not
hold. Therefore, with the required ordering of the roots, the no strut
condition cannot hold. A special case of this result is $a=0$,
ordinary Einstein-Maxwell theory, for which it is well known that one
cannot remove the conical singularities of the C-metric.

For $a^2>1$, $r_-\le 0<r_+$, the ordering of the roots reverts to
\be
u_1 < u_2 < u_3 < u_4=-{1\over Ar_-}
\ee
and a similar argument can be used to show that the no strut condition
cannot hold. In fact the argument is identical to the argument used in
E$\overline{\rm M}$D theory for $a^2>1$ above.

The special case $a^2=1$ requires more careful consideration. As in
the last section, when we were considering EM$\overline{\rm D}$ black
holes, we set $r_-=|\Sigma|(1-a^2)$ and take the limit $a^2\rightarrow
1$. This gives
\ba
F(u) &=& e^{-2|\Sigma|Au}, \nonumber \\
G(u) &=& \left[1-u^2(1+r_+Au)\right]e^{2|\Sigma|Au}.
\ea
Thus the fourth root $u_4$ of $G(u)$ disappears and $G(u)$ just has
the 3 roots $u_1<u_2<u_3$. The no strut condition
$G^\prime(u_2)+G^\prime(u_3)=0$ becomes
\be
(u_2-u_1)e^{2|\Sigma|Au_2} = (u_3-u_1)e^{2|\Sigma|Au_3}.
\ee
Since $u_2$ and $u_3$ are larger than $u_1$ and we are assuming here
that $A>0$, this condition may only be satisfied in the degenerate
case $u_2=u_3$.

Thus we see that, as in ordinary EMD theory, in EM$\overline{\rm D}$
theory one can {\em never} remove the conical singularities in the
C-metric to obtain a regular solution describing a pair of zero rest
mass black holes accelerating away from one another. This is as
expected, since the theory did not allow static zero rest mass black
holes of this type and so, although we expect the vacuum to be
unstable due to the presence of a scalar with negative kinetic energy,
the instability must manifest itself in a different way. One
possibility is that there may be a semi-classical instability leading
to the production of transparent massless wormholes of the type
discussed in section~2.2. This will be discussed in more detail in
section~5.

\subsection{E$\overline{\bf MD}$ Accelerating Solutions}

These solutions may be obtained from the previous solutions by
analytic continuation, setting the charge to be pure imaginary. The
metric is unchanged but the Maxwell potential becomes
\be
A = \sqrt{r_+r_-\over a^2-1}xd\phi.
\ee
Therefore, for $a^2<1$, $r_-\le 0$ and the proof that the no strut
condition can be satisfied is identical to the E$\overline{\rm M}$D
case. For $a^2>1$, $0\le r_-<r_+$ and it is easy to show that the
conical singularities can still be removed by an appropriate choice of
$r_+$ and $r_-$. In the special case $a^2=1$, we set
$r_-=|\Sigma|(a^2-1)$ and take the limit $a^2\rightarrow 1$ to obtain
\ba
F(u) &=& e^{2|\Sigma|Au}, \nonumber \\
G(u) &=& \left[1-u^2(1+r_+Au)\right]e^{-2|\Sigma|Au}.
\ea
The no strut condition then becomes
\be
(u_2-u_1)e^{-2|\Sigma|Au_2} = (u_3-u_1)e^{-2|\Sigma|Au_3}.
\ee
It is easy to see that this does now have solutions satisfying
$u_1<u_2<u_3$.

Thus, for all values of the dilaton coupling $a$ in E$\overline{\rm
MD}$ theory, it is possible to remove the conical singularities from
the C-metric to obtain a regular solution describing 2 zero rest mass
black holes with equal and opposite charges accelerating away from one
another. This is consistent with the result of the last section that,
for all $a$, E$\overline{\rm MD}$ theory allows static zero rest mass
black holes which are non-singular outside a regular event horizon.

\sect{Actions and Thermodynamics}

In this section we study the thermodynamic properties of the black
holes discussed in the last section. We do this first by calculating
their classical Euclidean actions since this will also be useful in
the later section on vacuum stability. Throughout this section, we
will assume that the scalar field $\sigma$ has positive kinetic
energy. We will work with the EMD action, since the E$\overline{\rm
M}$D solutions may be obtained simply by changing the sign of
$r_-$. The Euclidean action is given by
\be
I_E = -{1\over 16\pi}\int_{\cal M}d^4x\sqrt{g} \left\{R -
2(\partial\sigma)^2 - e^{-2a\sigma}F^2\right\} -{1\over
8\pi}\int_{\partial\cal M}d^3x\sqrt{h} \left[K\right]
\ee
where the boundary term is added to cancel second derivatives of the
metric appearing in the action and it ensures that $I_E$ is additive
over adjacent spacetime regions. The Euclidean dilaton black hole
solutions may be obtained from their Lorentzian counterparts by
setting $t=i\tau$~:
$$
ds^2 = \left(1-{r_+\over r}\right)\left(1-{r_-\over
r}\right)^{1-a^2\over 1+a^2}d\tau^2 + \left(1-{r_+\over
r}\right)^{-1}\left(1-{r_-\over r}\right)^{a^2-1\over 1+a^2}dr^2
$$
\be
+ r^2\left(1-{r_-\over r}\right)^{2a^2\over
1+a^2}(d\theta^2+\sin^2\theta d\phi^2).
\ee
Restricting ourselves to electrically charged solutions the
electromagnetic field strength 2-form becomes
\be
F = {i\over r^2}\sqrt{r_+r_-\over 1+a^2}d\tau\wedge dr.
\ee
In the EMD case we have $0\le r_-<r_+$ so that $F$ is pure imaginary
for the Euclidean solutions. In E$\overline{\rm M}$D theory we must
replace $r_-$ by $-r_-$ in $F$ but now $r_-\le 0<r_+$ so that $F$ is
still pure imaginary.

Note that the metric is apparently singular at $r=r_+$. To examine
the metric just outside $r=r_+$ more closely, in the non-extreme case,
set $r=r_++\varepsilon^2$ and look at the line element for small
$\varepsilon$~:
\be
ds^2 \approx {\varepsilon^2\over r_+}\left(1-{r_-\over
r_+}\right)^{1-a^2\over 1+a^2}d\tau^2 + 4r_+\left(1-{r_-\over
r_+}\right)^{a^2-1\over 1+a^2}d\varepsilon^2,
\ee
neglecting the $\theta,\phi$ dependance. Clearly $r=r_+$
($\varepsilon=0$) is a conical singularity which may be removed by
identifying $\tau$ with period $\beta$ where
\be
\beta = 4\pi r_+\left(1-{r_-\over r_+}\right)^{a^2-1\over 1+a^2}.
\label{beta}
\ee

\subsection{Actions for Non-Extreme Black Holes}

The calculation of the action is simplified by making use of the
Einstein equations which imply that for on-shell solutions
$R=2(\partial\sigma)^2$ and so the action reduces to
\be
I_E = {1\over 16\pi}\int_{\cal M}d^4x\sqrt{g}e^{-2a\sigma}F^2 -{1\over
 8\pi}\int_{\partial\cal M}d^3x\sqrt{h} \left[K\right].
\ee
The volume term may be evaluated very simply. We have spherical
symmetry and $\tau\in[0,\beta]$, $r\in[r_+,\infty)$ so, absorbing the
dilaton factor into the integration measure, the integral reduces to
\be
\int_{\cal M}d^4x\sqrt{g}e^{-2a\sigma}\cdots =
4\pi\beta\int_{r_+}^\infty dr~r^2\cdots.
\ee
The scalar invariant $F^2$ is given by
\be
F^2 = - {2r_+r_-\over(1+a^2)r^4}
\ee
and so the volume integral contribution to the action is
\be
{1\over 16\pi}\int_{\cal M}d^4x\sqrt{g}e^{-2a\sigma}F^2 = - {\beta
r_-\over 2(1+a^2)}.
\label{Vol}
\ee
In the E$\overline{\rm M}$D case, $F^2$ becomes
$+{2r_+r_-\over(1+a^2)r^4}$, with $r_-$ negative, but the sign of the
$F^2$ term in the action is changed and so the contribution to the
action is given by the same formula (\ref{Vol}).

To evaluate the boundary term, it is necessary to identify precisely
where the boundaries are. It is here that the extreme solutions differ
from the non-extreme solutions. We will deal with the non-extreme case
first. In this case, the topology of the Euclidean solution is the
well-known cigar shape, see Fig.~1 ($\theta,\phi$ coordinates
suppressed).
\begin{figure}
\epsffile{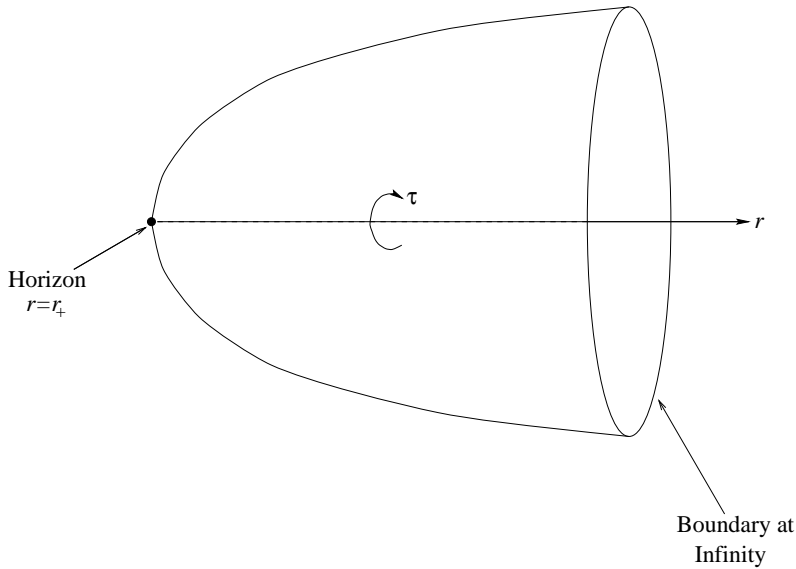}
\caption{Geometry of non-extreme Euclidean dilaton black holes}
\end{figure}
Thus the only boundary is at $r=\infty$. The contribution to the
action is the integral over this boundary of $\left[K\right]$, the
trace of the extrinsic curvature of the boundary minus a term to
ensure that the action of flat space is zero. Shifting the radial
coordinate $r\rightarrow r-{a^2r_-\over 1+a^2}$ so that, for large
$r$, a sphere of radius $r$ has surface area $4\pi r^2+{\cal O}(1)$,
the metric for large $r$ becomes
\be
ds^2 \sim \left(1-{2M\over r}\right)d\tau^2 + \left(1+{2M\over
r}\right)dr^2 + r^2(d\theta^2+\sin^2\theta d\phi^2)
\ee
where
\be
M = {1\over 2}\left(r_++{1-a^2\over 1+a^2}r_-\right).
\ee
The normal to the surface $r=R$ is $n_\mu=(0,N,0,0)$ where
$N=\sqrt{g_{rr}}=1+{M\over R}+{\cal O}(R^{-2})$ is the lapse
function. The trace of the extrinsic curvature is
\ba
K &=& {1\over 2N}\left[g^{\tau\tau}g_{\tau\tau,r} +
g^{\theta\theta}g_{\theta\theta,r} + g^{\phi\phi}g_{\phi\phi,r}\right]
\nonumber \\
 &=& {2\over R} - {M\over R^2} + {\cal O}(R^{-3}).
\ea
Thus we may identify the $2\over R$ term as the contribution from flat
${\Bbb R}^4$ which must be subtracted to ensure that the action of
flat space is zero. This gives $\left[K\right]=-{M\over R^2}+{\cal
O}(R^{-3})$. The surface integral over the surface $r=R$ is
\be
\int_{r=R}d^3x\sqrt{h} = 4\pi\beta R^2 + {\cal O}(R).
\ee
Therefore the boundary integral contribution to the action is
\be
-{1\over 8\pi}\int_{\partial\cal M}d^3x\sqrt{h} \left[K\right] =
 \half\beta M.
\ee
Thus the total Euclidean action may be written as
\be
I_E = \fraction{1}{4}\beta(r_+-r_-) = \half\beta(M-Q\Phi)
\ee
where $\Phi={Q\over r_+}$ is the electric potential on the horizon,
in a gauge in which $A_\tau$ vanishes at infinity. Using the
expression for $\beta$ (\ref{beta}) we may write $I_E$ in terms of
$r_\pm$~:
\be
I_E = \pi r_+^2\left(1-{r_-\over r_+}\right)^{2a^2\over 1+a^2}.
\ee
This is precisely $\fraction{1}{4}{\cal A}$ where $\cal A$ is the area
of the black hole event horizon. This formula, in terms of $r_\pm$,
continues to be true for E$\overline{\rm M}$D black holes.

\subsection{Thermodynamics}

In ordinary Einstein-Maxwell theory the \RN\ black hole is expected to
undergo Hawking radiation and thereby lose mass. Assuming that the
black hole emits neutral particles, the black hole charge will remain
constant and the mass will decrease. Eventually the black hole will
approach extremality and the Hawking radiation will turn off. It is
well known that this process takes an infinitely long time and so an
extreme \RN\ solution is never actually produced. In the EMD case, the
situation is rather different \cite{GibMae}. In particular, for
$a^2>1$ the Hawking temperature blows up as one approaches the extreme
solution. Also, however, the area of the event horizon decreases to
zero. It is thus interesting to ask which effect dominates and how
long it takes to produce an extreme solution.

In the E$\overline{\rm M}$D case there are no extreme solutions and
the Hawking radiation is never turned off. One is then lead naturally
to ask what happens as such black holes evaporate. In the case $a^2<1$
when regular black holes with $M\le 0$ are allowed, there seems to be
no reason why the black holes cannot evaporate away {\em all} of their
mass. But what happens next? Is there some lower bound on the mass or
will it continue to decrease to $-\infty$? How long will this process
take? In the case $a^2\ge 1$ there {\em is} a lower bound on the mass,
$M\ge|Q|\sqrt{a^2-1}$, and again there are no extreme solutions. So
what is the endpoint of the evaporation process in this case?

In the study of thermodynamics it is useful to define thermodynamic
potentials. Let us define the Gibbs Free Energy~:
\be
W = M - TS - Q\Phi = -T\log Z.
\ee
In the semi-classical approximation we approximate the partition
function $Z$ by $e^{-I_E}$. Thus we have
\be
W = TI_E = \half(M-Q\Phi).
\ee
This then gives the well known result
\be
S = I_E = \fraction{1}{4}{\cal A}
\ee
and also the Smarr formula for the mass~:
\be
M = 2TS + Q\Phi.
\ee
Regarding $M$ as a homogeneous function of $S$ and $Q$ we obtain
directly the First Law of black hole thermodynamics~:
\be
dM = TdS + \Phi dQ.
\ee

Since the evaporation is via neutral particles, we must consider
processes taking place at constant $Q$. For example, consider the
specific heat at constant $Q$~:
\be
C_Q = \left.{\partial M\over \partial T}\right|_Q = T\left.{\partial
S\over \partial T}\right|_Q.
\ee
In terms of $r_\pm$ this is
\be
C_Q = -2\pi r_+^2\left(1-{1-a^2\over 1+a^2}{r_-\over
r_+}\right)\left(1-{3-a^2\over 1+a^2}{r_-\over
r_+}\right)^{-1}\left(1-{r_-\over r_+}\right)^{2a^2\over 1+a^2}.
\label{SpHeat}
\ee
In ordinary EMD theory we have $0\le r_-<r_+$. For $a^2\ne 1$ one may
express $C_Q\over Q^2$ as a function of $M\over Q$, see Fig.~2.
\begin{figure}
\epsffile{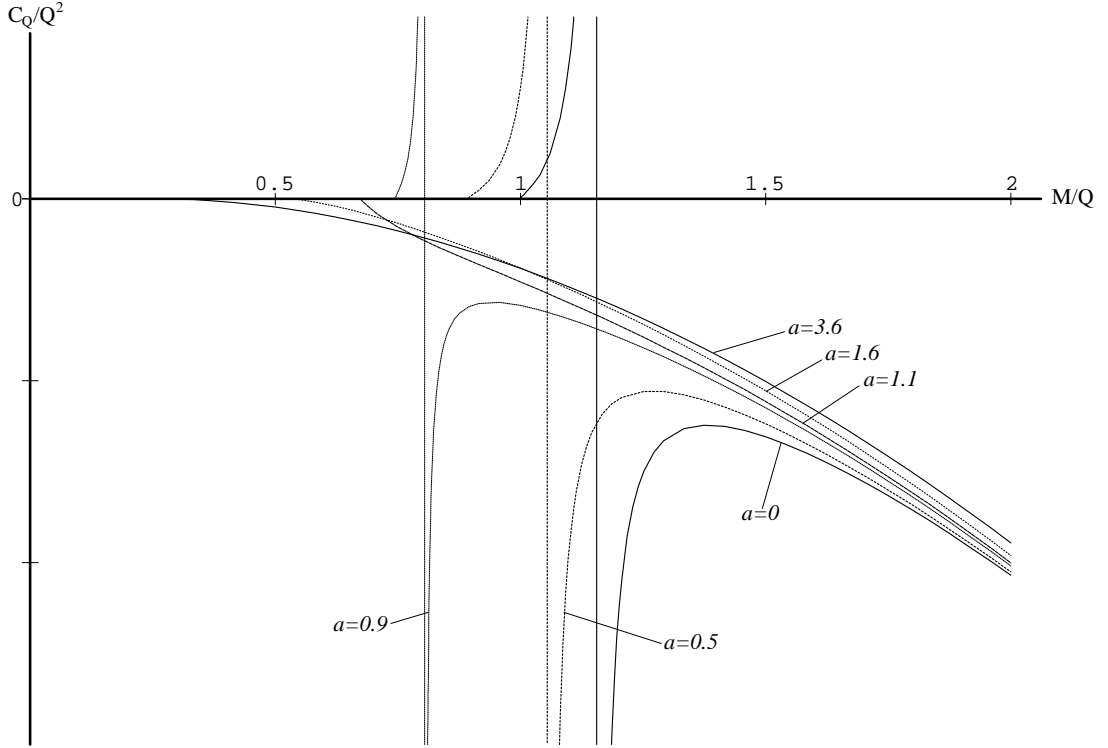}
\caption{Specific heats of EMD black holes for various values of $a$}
\end{figure}
For $a^2<1$ the specific heat diverges when
\be
r_+ = {3-a^2\over 1+a^2}r_- \qquad\mbox{i.e.}\qquad \left|{M\over
Q}\right| = {2-a^2\over\sqrt{3-a^2}}.
\ee
For $a^2>1$ this divergence is excluded by the requirement that
$r_+>r_-$. The origin of this divergence can be seen by considering
behaviour of the temperature as $M$ varies. Initially, for large $M$,
the temperature increases as $M$ decreases. However, for $a^2<1$, the
temperature ultimately approaches zero as one approaches extremality
and so there must be a turning point in the temperature as a function
of $M$ at constant $Q$. This then is the divergence of the specific
heat $C_Q$. In the case $a^2>1$, however, the temperature diverges as
one approaches extremality and so there is no such turning point. The
case $a^2=1$ is somewhat special. In this case the specific heat
reduces to
\be
C_Q = -2\pi r_+^2 = -8\pi M^2
\ee
{\em independent of} $Q$. The temperature is ${1\over 8\pi M}$, also
independent of $Q$. This is the case which arises in string theory and
we have the interesting result that the thermodynamic properties at
constant charge are {\em identical} to those of the the Schwarzschild
solution in General Relativity. This may be seen as a result of the
fact that for $a^2=1$ the metric is the same as the Schwarzschild
solution except for the angular components of the metric which do not
affect its thermodynamic properties. We may thus deduce immediately
that the black hole will radiate away its mass faster and faster,
reaching extremality in a {\em finite} time.

To estimate the time taken for the more general EMD black holes to
radiate sufficient mass to become extreme, we use Stefan's law to
approximate the rate of loss of mass of the black hole~:
\be
{dM\over dt} \approx -\sigma{\cal A}T^4
\ee
which gives
\be
{dM\over dt} \propto {1\over r_+^2}\left(1-{r_-\over
r_+}\right)^{4-2a^2\over 1+a^2}
\ee
with $Q$ held constant. For nearly extreme solutions, we may set
$M={Q\over\sqrt{1+a^2}}+\varepsilon$ giving
\be
{d\varepsilon\over dt} \propto \varepsilon^{4-2a^2\over 1+a^2}.
\ee
Thus the time taken to reach extremality will be infinite for $a^2<1$
and finite for $a^2\ge 1$. Since the area of the event horizon
vanishes for extreme EMD black holes, this semi-classical description
of the black hole thermodynamics is expected to break down near
extremality and a more complete quantum theory is needed to make
further predictions about what happens near the end of the
evaporation.

In E$\overline{\rm M}$D theory $r_-\le 0<r_+$ and there are no extreme
solutions. So what is the endpoint of the evaporation? The specific
heat at constant charge is again given by (\ref{SpHeat}) but now
$r_-\le 0$ and the electric charge is given by $Q=\sqrt{-r_+r_-\over
1+a^2}$. For $a^2\ne 1$ the behaviour of $C_Q\over Q^2$ as a function
of $M/Q$ is shown in Fig.~3.
\begin{figure}
\epsffile{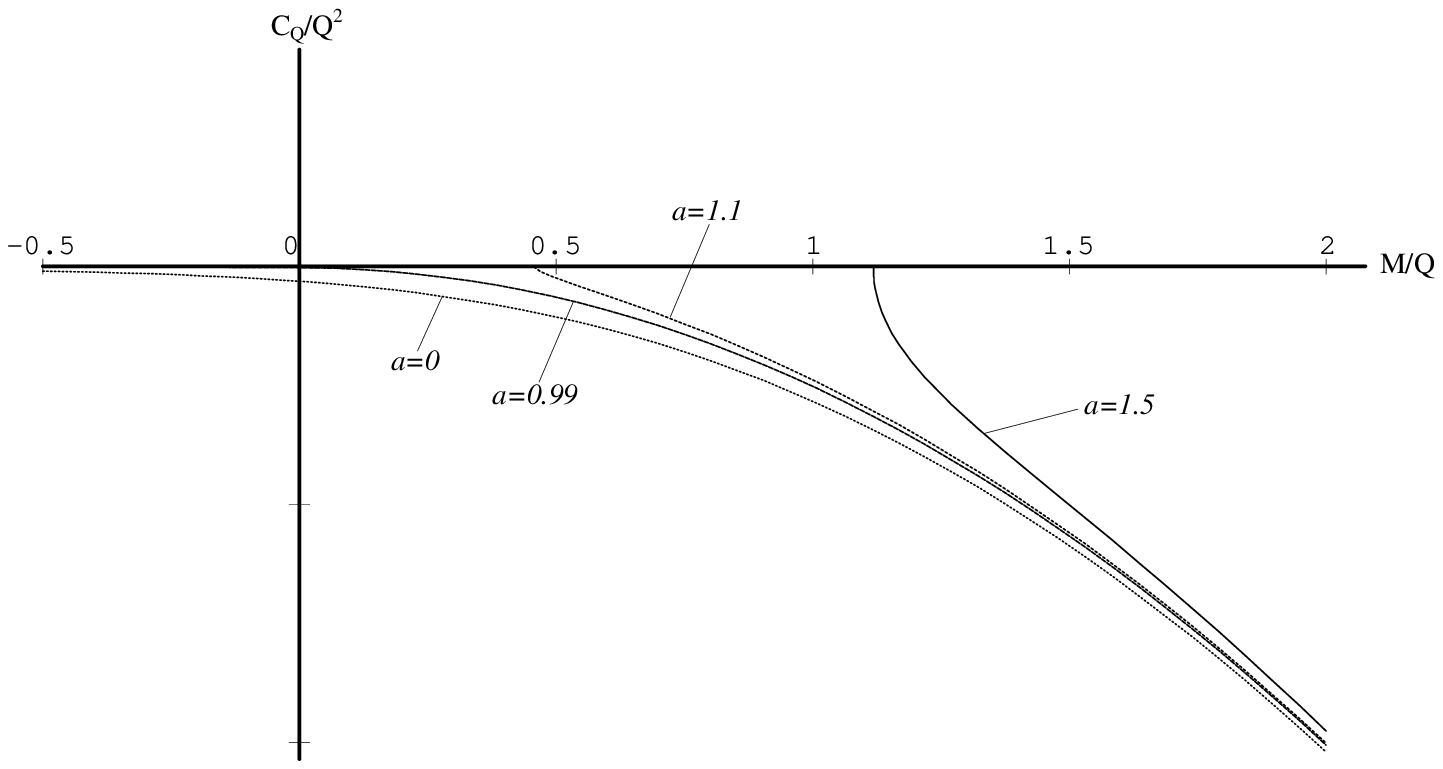}
\caption{Specific heats of E$\overline{\rm M}$D black holes for
various values of $a$}
\end{figure}
Again the behaviour is qualitatively different depending on whether
$a^2$ is greater than or less than 1. For $a^2<1$ there is no lower
bound on the mass and the specific heat approaches zero as
$M\rightarrow -\infty$. This can be seen as a result of the fact that
the temperature $T$ increases without bound as $M\rightarrow -\infty$
with $Q$ held constant (see Fig.~4). A natural question to ask is then
\begin{figure}
\epsffile{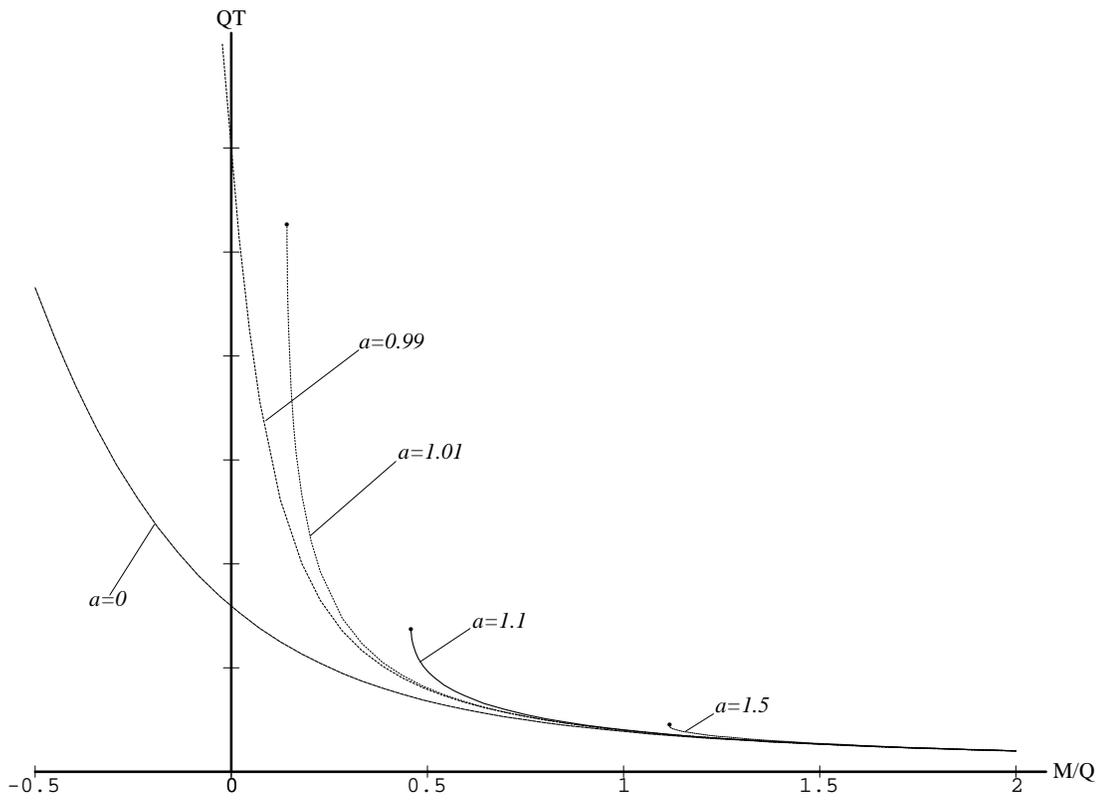}
\caption{Temperatures of E$\overline{\rm M}$D black holes for
various values of $a$}
\end{figure}
how quickly will the black hole radiate away mass and how long will it
take for $M$ to reach $-\infty$. As ${M\over|Q|}\rightarrow -\infty$,
the outer horizon shrinks as $r_+\sim{1\over|M|}$ and the inner
horizon approaches $-\infty$ like $M$. Thus the temperature $T$
diverges like
\be
T \sim |M|^{3-a^2\over 1+a^2}
\ee
and the area of the event horizon decreases to zero like
\be
{\cal A} \sim |M|^{-{4\over 1+a^2}}.
\ee
Thus, if we approximate the rate of loss of mass by Stefan's Law, we
have
\be
{dM\over dt} \sim -|M|^{8-4a^2\over 1+a^2}.
\ee
Therefore, since $a^2<1$, this rather crude approximation for the
evaporation rate tells us that the black hole's mass will reach
$-\infty$ within a {\em finite} time. Of course, as $M\rightarrow
-\infty$, the event horizon shrinks to zero size and so this
semi-classical description will break down. Nonetheless this simple
calculation is yet another indicator of the instability of these black
holes.

For $a^2>1$ the situation is rather different. In this case the mass
is bounded below~:~$M\ge|Q|\sqrt{a^2-1}$ and no static spherically
symmetric solutions exist with masses smaller than this. Also, as was
pointed out in section~2, solutions saturating this bound are {\em
not} extreme. They have $r_+\ne r_-$ and $T\ne 0$. Fig.~4 shows that
as $M\rightarrow|Q|\sqrt{a^2-1}$ the temperature $T$ becomes finite
and non-zero. Also the specific heat $C_Q$ vanishes in this
limit. How long does a black hole take to radiate away sufficient mass
to saturate this bound? What happens to black holes which do saturate
this mass bound? Since the temperature $T$ and event horizon area
$\cal A$ both remain finite and non-zero in this limit, Stefan's law
predicts that E$\overline{\rm M}$D black holes with $a^2>1$ should
evaporate and reach $M=|Q|\sqrt{a^2-1}$ within a {\em finite}
time. Once the black hole reaches this limit it can no longer lose
mass at constant charge and remain static and spherically symmetric
since no such solutions exist with $M<|Q|\sqrt{a^2-1}$. Of course,
when {\em any} black hole loses mass by the emission of a particle,
the spacetime during the process is neither static nor spherically
symmetric. However, in the case of a stable black hole such as the
Schwarzschild solution, the spacetime may settle down to a new
Schwarzschild solution with reduced mass after the emission of the
particle. In the present case, however, there is no such static
spherically symmetric solution for the spacetime to settle down to. We
may thus expect some sort of catastrophic quantum mechanical
instability to be exhibited by these black holes.

The case $a^2=1$ is again somewhat special. The temperature and
specific heat are given by
\be
T = {1\over 8\pi M} \quad\mbox{and}\quad C_Q = -8\pi M^2,
\ee
{\em independent of $Q$}. Note that once again these thermodynamic
quantities are identical to those of the Schwarzschild solution. Also,
as for the Schwarzschild solution, the mass is positive,
$M>0$. However, unlike the Schwarzschild solution, these black holes
have electric charge and so we cannot set $M=0$ unless $Q=0$. If it
were the case that these black holes with $a^2=1$ only emitted neutral
particles then, as for the Schwarzschild black hole in the
semi-classical approximation, they would radiate away all of their
mass within a finite time. This is not possible, however, since no
solution exists with $M=0$ and $Q\ne 0$. This problem is easily
resolved because as $M$ decreases with $Q$ held fixed, the ratio
$Q\over M$ increases without bound, as does the electric potential
$\Phi$ on the horizon. Eventually this will become large enough to
give a significant probability for the emission of charged particles
by the black hole. Thus the black hole may radiate away all of its
charge as well as all of its mass.

\subsection{Actions for Extreme Black Holes}

To calculate the Euclidean action for extreme dilaton black holes
($r_+=r_-=r_H$) we need to take into account the fact that the
topology of these solutions is different from that of non-extreme
solutions and so there is potentially an extra contribution to the
action from the inner boundary. The Euclidean metric is
$$
ds^2 = \left(1-{r_H\over r}\right)^{2\over 1+a^2}d\tau^2 +
\left(1-{r_H\over r}\right)^{-{2\over 1+a^2}}dr^2
$$
\be
+ r^2\left(1-{r_H\over r}\right)^{2a^2\over
1+a^2}(d\theta^2+\sin^2\theta d\phi^2).
\ee
There are 2 distinct cases to be considered~: $a=0$ and $a\ne 0$. If
$a=0$ we have the extreme electric Euclidean \RN\ solution. The
horizon $r=r_H$ is infinitely far away along any curve in this space
and so the topology is the the well-known pipette shape, see Fig.~5.
\begin{figure}
\epsffile{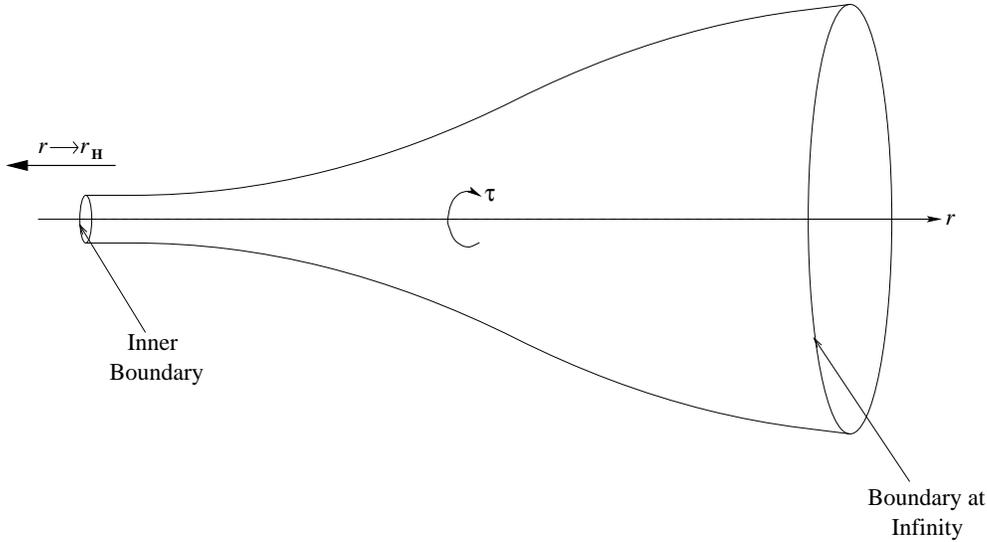}
\caption{Geometry of extreme Euclidean \RN\ black hole}
\end{figure}
Since the point $r=r_H$ is no longer a point in the space, we must add
to the action a contribution from the extrinsic curvature integrated
over an inner boundary just outside $r=r_H$. If $a\ne 0$ the point
$r=r_H$ is a curvature singularity. We may deal with this by cutting
it out of the space by introducing a boundary just outside $r=r_H$. In
either case there is potentially a contribution to the Euclidean
action due to the integral of the extrinsic curvature over this
boundary. However, it is easy to check that this boundary term
vanishes as the inner boundary is taken to $r_H$.

In the non-extreme case we were forced to identify $\tau$ with a
particular period $\beta$ in order to remove a conical singularity on
the horizon. In the extreme case, the horizon is no longer a part of
the space and so we are free to identify $\tau$ with any period
$\beta$ we choose, or not at all. Clearly then the Euclidean action
vanishes in the extreme limit $r_+\rightarrow r_-$ since the
contributions from the boundary term at infinity $\half\beta M$ and
from the volume integral of $F^2$ cancel. Note that this would not be
the case if we considered magnetically charged black holes since in
that case $F$ would be real and the volume integral would give a {\em
positive} contribution of $\half\beta M$ to the action. The Euclidean
action would then be $I_E=\beta M$.

\sect{Massless Wormholes}

If one considers a scalar field $\sigma$ with negative kinetic energy
rather than a vector field then regular massless ultra-static wormhole
solutions are possible. In four spacetime dimensions the transparent
massless wormhole metric (\ref{Worm}), in isotropic coordinates, is
given by
\be
ds^2 = -dt^2 + {c^2\over 4}\left(1+{1\over
r^2}\right)^2(dx^2+dy^2+dz^2),
\ee
with $r^2 = x^2 + y^2 +z^2$ and the scalar field is \be \sigma =
\tan^{-1}\left(r^2-1\over 2r\right), \ee where $c$ is a constant.

The spatial sections have the form of an Einstein-Rosen bridge joining
two isometric regions each with vanishing ADM mass. The isometry
interchanging the regions is the inversion~:
\be
r \rightarrow {1\over r}.
\ee

Because the metric is ultra-static (i.e.\ $g_{00}=-1$) there is no
horizon hiding the two sides of the bridge from one another as there
is for black holes. Moreover the Newtonian gravitational potential
$U=-{1\over 2}\ln(-g_{00})$ is constant and therefore the wormhole
exerts no gravitational attraction.

Because the $g_{00}=-1$ we may regard the solution is an instanton of
a three-dimensional theory. The four-dimensional Einstein equations
are
\be
R_{\mu\nu} = -2(\partial_\mu\sigma)(\partial_\nu\sigma)
\ee
where $\mu, \nu=0,1,2,3$. The equations are trivially satisfied if
either $\mu$ or $\nu=0$. For $\mu, \nu=1,2,3,$ they give the
Euler-Lagrange equations of a Euclidean three-dimensional theory with
action (modulo boundary terms)
\be
{1\over 16\pi} \int d^3x\sqrt{g} \left\{R +
2(\partial\sigma)^2\right\}
\ee
in which the field $\sigma$ contributes {\em negatively} to the
action. This unusual sign typically arises in wormhole theories when
one analytically continues a Lorentzian theory (in the present case a
(2+1)-dimensional theory) containing a pseudo-scalar field to a
Riemannian one.  An alternative viewpoint is to use Hodge duality to
replace $d\sigma$ by a two-form
\be
 d\sigma = \star F.
\ee
One then gets a theory of gravity coupled to abelian electrodynamics.
In the absence of gravity it has been pointed out by Polyakov
\cite{Pol} that this theory admits (singular) instantons which
correspond to Dirac monopoles. It is interesting that if one couples
this system to gravity (which has no dynamical degrees of freedom in
three dimensions) these Polyakov instantons become non-singular.

In theories without gravity one often says that non-singular
instantons in $d$ spatial dimensions may be interpreted as
non-singular solitons in $d+1$ spacetime dimensions. The present
example with $d=3$, and its obvious generalization to higher $d$ show
clearly that if gravity is involved then the soliton may be rather
exotic. In particular it may be massless. Note, however, that if one
really were considering gravity coupled to abelian electrodynamics,
i.e.\ Einstein-Maxwell theory rather than gravity coupled to a scalar
field, then the three-dimensional solutions {\em cannot} be trivially
promoted to four-dimensional solutions. The field equations in four
dimensions are
\be
R_{\alpha\beta} - \half Rg_{\alpha\beta} =
2F_{\alpha\mu}F_{\beta\nu}g^{\mu\nu} - \half
g_{\alpha\beta}F_{\mu\nu}F^{\mu\nu}.
\ee
The spatial equations agree with what one would get in three
dimensions but the $0-0$ equation imposes an extra constraint which
the Einstein-Polyakov instantons do not satisfy. This must be so
because we know that Einstein-Maxwell theory satisfies the usual
positive and dominant energy conditions and does not admit
non-singular solutions with zero ADM mass.

It seems reasonable to conjecture that if the zero mass wormhole
solutions were perturbed they would become singular and/or accelerate
off to infinity. It would be interesting to know therefore whether
there exist non-singular accelerating solutions.  The obvious
procedures seem only to lead accelerating solutions with
singularities. It would also be interesting to know whether
multi-solutions exist representing more than one wormhole at rest.

Another question relates to ``tachyonic'' solutions. Tachyons should
be thought of as representing what happens to an unstable background
after the instability has set in rather than indicating any violation
of causality. It is clear that the characteristics of a scalar field
are still given by the metric regardless of whether it contributes
negatively to the energy of the system.
  
Tachyonic solutions will be discussed in greater detail later. For the
time being we note that they should be invariant under ${\Bbb R}\times
SO(2,1)$ or $SO(2)\times SO(2,1)$ rather than ${\Bbb R}\times
SO(3)$. To obtain them in the present case one makes the replacement
$t\rightarrow iz$ and $z\rightarrow it$ with $z$ and $t$ real and thus
$r^2\rightarrow x^2+y^2-t^2$. The resulting solutions
\be
ds^2 = dz^2 + {c^2\over 4}\left(1+{1\over
r^2}\right)^2(dx^2+dy^2-dt^2)
\ee
are invariant under translations in the new $z$ coordinate and are
nakedly singular on the Cerenkov cone given by
\be
x^2+y^2 = t^2.
\ee
Note that if we set $t=0$ we obtain time-symmetric initial data on
${\Bbb R}\times ({\Bbb R}^2-\{0\})$, i.e.\ the product of a
two-dimensional wormhole with the real line. The three-metric
\be
ds^2 = dz^2 + {c^2\over 4}\left(1+ {1\over
x^2+y^2}\right)^2(dx^2+dy^2)
\ee
is non-singular but has zero ADM mass. The resulting spacetime
subsequently becomes nakedly singular on the Cerenkov cone.

This behaviour is similar to that of $\lambda\phi^{2n\over n-2}$
scalar field theory in $n$ flat spacetime dimensions where the sign of
$\lambda$ is such as to give a negative potential energy. There are
solutions of the form
\be
1\over(a^2+{\bf x}^2-t^2)^{n-2\over 2}. 
\ee
The solution has non singular time-symmetric initial data of finite
energy at $t=0$ but blows up on the spacelike hyperboloid
\be
t = \sqrt{a^2+{\bf x}^2}.
\ee

\sect{Tachyonic solutions} 

We saw in the previous section the relevance of $SO(2,1)$ invariant
solutions for the instability process. We shall refer to these as
``tachyonic solutions'' with the same understanding as before -- they
do not signal acausality but rather instability.

In vacuum or Einstein-Maxwell theory assuming that $SO(2,1)$ acts on
two-dimensional orbits implies, by a simple generalization of
Birkhoff's theorem, the existence of an additional spacelike Killing
vector.  If this had non-compact orbits the spacetime would then have
the same symmetry group as that of a spacelike world line in flat
Minkowski spacetime. However there is a surprise. The replacement
\ba
\theta &\rightarrow & {\pi\over 2} + it \nonumber \\
t &\rightarrow & iZ.
\ea
in the standard Schwarzschild metric leads to
\be
ds^2 = \left(1-{r_+\over r}\right)dZ^2 + \left(1-{r_+\over
r}\right)^{-1}dr^2 + r^2(-dt^2+\cosh^2td\phi^2).
\label{TachS}
\ee

This metric is indeed invariant under translations along the direction
of motion of the ``tachyon'' (i.e.\ in the $Z$ direction) and reversal
of the $Z$ coordinate. In fact the symmetries of the metric become
more transparent if one introduces pseudo isotropic coordinates
$T,X,Y$ by
\ba
T &=& s\sinh t \nonumber \\
X &=& s\cosh t \cos\phi \\
Y &=& s\cosh t \sin\phi \nonumber
\ea
with $r=s(1+{r_+\over 4s})^2$. The metric then becomes
\be
ds^2 = \left({1-{r_+\over 4s} \over 1+{r_+\over 4s}}\right)^2 d Z^2 +
\left(1+{r_+\over 4s}\right)^4(-dT^2+d X^2+dY^2).
\ee

This metric has been interpreted as that of a tachyon, the surface
$r=2M$ being thought of as the analogue of a Cerenkov cone \cite{Per}.
However this interpretation is not really tenable because if $r_+$ is
taken to be positive then the metric is complete and everywhere
non-singular only as long as the spatial coordinate $Z$ is identified
modulo $4\pi r_+$~:
\be
0 \le Z \le 4\pi r_+.
\ee
The variable $s$ then runs from $s=\fraction{1}{4}r_+$ to infinity and
the pseudo isotropic coordinates $(T,Y,Y)$ are constrained to lie
outside the hyperboloid
\be
X^2 + Y^2 - T^2 \ge {r_+^2\over 4}.
\ee
The Killing horizon at $r=r_+$ (i.e.\ $s={1\over 4}r_+$) is a null
surface. One might be tempted to say that in some sense the presence
of the tachyon has brought about the `compactification' of space
and the restriction of spacetime to the exterior of its Cerenkov
cone.

In the light of our previous discussion in section 4 and the closely
related case of the instability of the Kaluza-Klein vacuum \cite{Wit}
a more satisfactory interpretation is to regard this Lorentzian metric
as the result of a tunnelling instability of the flat spacetime on
${\Bbb R}^3\times S^1$ where the $S^1$ factor refers to the periodic
spacelike coordinate $Z$. Alternatively one may think of it as
providing the Cauchy development of the non-singular time symmetric
initial data set on $S^1\times {\Bbb R}^2$ obtained by putting $t=0$
in (\ref{TachS}).

\sect{Semi-Classical Vacuum Decay}

The existence of negative energy, asymptotically Minkowskian black
hole solutions in Einstein-anti-Maxwell theory is a strong indication
that the vacuum of the theory is unstable. Further evidence for this
instability was provided in the sections describing accelerating black
holes and their thermodynamics. A more convincing demonstration of the
instability of the vacuum would be provided by an instanton solution
describing its decay. We therefore look for a Euclidean solution which
approaches the vacuum at infinity, i.e.\ it must be asymptotically
${\Bbb R}^4$.

As we saw in the last section, the Euclidean Schwarzschild solution
will not do because, to avoid a singularity at the horizon $r=r_+$, it
was necessary to periodically identify the former time coordinate
$Z$. The solution is thus asymptotically ${\Bbb R}^3\times S^1$. Note
that the period of this identification of the imaginary time
coordinate, $\beta=4\pi r_+$, is precisely $1\over T$ where $T$ is the
Hawking temperature of the Lorentzian black hole solution. This
suggests that we should consider the extreme \RN\ solution since this
has $T=0$ and so we would be free to identify $Z$ with whatever period
we liked (or not at all). There is a slight complication in that, in
order to obtain the Euclidean \RN\ solution, we need to take $Q$ to be
pure imaginary. However, since this is precisely what we had to do to
obtain the anti-\RN\ solution, we will obtain the desired solution of
the Euclidean Einstein-anti-Maxwell equations with $Q$ real. To see
this in more detail, consider the $Q=M$ Lorentzian anti-\RN\
solution~:
$$
ds^2 = -\left(1-{2M\over r}-{Q^2\over r^2}\right)dt^2 +
\left(1-{2M\over r}-{Q^2\over r^2}\right)^{-1}dr^2 +
r^2(d\theta^2+\sin^2\theta d\phi^2),
$$
\be
A = {Q\over r}dt.
\ee
Here $Q=M$ is real and the horizons satisfy $r_-<0<r_+$, so this is
{\em not} an extreme solution (in fact as we saw in section 2, there
are no extreme solutions in this theory). Now we obtain the Euclidean
solution by setting $t=iz$ and, in order to keep the vector potential
1-form $A$ real, we must also make the replacement $Q\rightarrow
iQ$. We thus obtain the following Euclidean solution of the
Einstein-anti-Maxwell equations
$$
ds^2 = \left(1-{M\over r}\right)^2dz^2 + \left(1-{M\over
r}\right)^{-2}dr^2 + r^2(d\theta^2+\sin^2\theta d\phi^2),
$$
\be
A = {M\over r}dz,
\label{Instant}
\ee
which is our desired instanton. In isotropic coordinates, $r=\rho+M$,
it may be written as
\be
ds^2 = \left(1+{M\over\rho}\right)^{-2}dz^2 +
\left(1+{M\over\rho}\right)^2\left[dx^2+dy^2+d\tau^2\right]
\ee
where $\rho^2=x^2+y^2+\tau^2$. Note that this space, restricted to
$\rho>0$, is everywhere non-singular and geodesically complete. The
singularity at $\rho=0$ is at an infinite proper distance along any
geodesic. There is no need to periodically identify $z$, or
alternatively, we are free to identify $z$ with any period we choose.

To see how the above instanton leads to the instability of the vacuum,
we analytically continue to the Lorentzian solution by setting
$\tau=it$~:
$$
ds^2 = \left(1+{M\over\rho}\right)^{-2}dz^2 +
\left(1+{M\over\rho}\right)^2\left[dx^2+dy^2-dt^2\right],
$$
\be
A = {M\over\rho+M}dz
\ee
where $\rho^2=x^2+y^2-t^2$. This then is the solution into which the
vacuum decays. It is a tachyon solution, as in the last section,
describing a charged particle moving with infinite speed along the
$z$-axis. However, here we wish to interpret it differently as
representing the instability of the vacuum in Einstein-anti-Maxwell
theory. It is still everywhere non-singular and geodesically complete
since the proper distance down the ``infinite throat'' at $\rho=0$ is
infinite along all geodesics. However, it is clear that this throat
region expands outwards radially in all directions perpendicular to
the $z$-axis at the speed of light.

Due to the cylindrical symmetry of the solution, it is best described
in cylindrical polar coordinates,
$r=\sqrt{x^2+y^2},\theta=\tan^{-1}{x\over y}$~:
\be
ds^2 = \left(1+{M\over\sqrt{r^2-t^2}}\right)^{-2}dz^2 +
\left(1+{M\over\sqrt{r^2-t^2}}\right)^2\left[dr^2+r^2d\theta^2-dt^2\right].
\ee
Thus, at t=0, an infinite throat forms along the $z$ axis. The
$z$-axis becomes infinitely far away along geodesics in the new
spacetime and this throat region expands away from the $z$-axis at the
speed of light. Actually it is this rapid expansion which is
responsible for the $z$-axis being infinitely far away. If one
considers a constant time slice, then it is easy to see that $r=0$ is
not at an infinite proper distance from points with $r>0$. However,
such curves in the constant time slice are not geodesics. After the
formation of the throat, the topology of the spacetime changes to
${\Bbb R}^{1,1}\times\left({\Bbb R}^2-\{0\}\right)$. The above
solution describes one such throat region forming and expanding to
infinity at the speed of light. It is more realistic, physically, to
suppose that actually such throats will form throughout the vacuum at
a given rate per unit volume and then expand outwards until they
collide with one another. Of course, an observer in the spacetime will
never see a throat form because, since it travels outwards at the
speed of light, by the time he sees it, he will have already fallen
down it! To investigate the properties of the spacetime ``down the
throat'', i.e.\ near $r=t$, define new coordinates $u,T$ by
\ba
r &=& e^u\cosh T \nonumber \\
t &=& e^u\sinh T
\ea
then the metric becomes
\be
ds^2 \approx {e^{2u}\over M^2}dz^2 + M^2du^2 - M^2dT^2 +
M^2\cosh^2Td\theta^2.
\ee
This is a completely non-singular, cylindrically symmetric spacetime
(${\Bbb R}^{1,2}\times S^1$) such that the radius of the $S^1$
increases rapidly as the time coordinate $T$ increases.

Finally, in order to show that the above mechanism for the decay of
the Einstein-anti-Maxwell vacuum is important, we need to calculate
the probability of the production of such infinite throats. According
to the semi-classical approximation, this probability is $P=e^{-I_E}$
where $I_E$ is the action of the Euclidean instanton
(\ref{Instant}). But (\ref{Instant}) is equivalent to the extreme
Euclidean \RN\ solution of the Euclidean Einstein-Maxwell equations
and so, as we saw in section~4.3, its action is zero. Therefore one
expects these infinite throats to be copiously produced and hence, as
expected, the Einstein-anti-Maxwell vacuum will be genuinely unstable.

A similar construction gives the decay process for the E$\overline{\rm
M}$D vacuum. We start from the E$\overline{\rm M}$D black hole
solutions of section~2~:
$$
ds^2 = -\left(1-{r_+\over r}\right)\left(1-{r_-\over
r}\right)^{1-a^2\over 1+a^2}dt^2 + \left(1-{r_+\over
r}\right)^{-1}\left(1-{r_-\over r}\right)^{a^2-1\over 1+a^2}dr^2
$$
\be
+ r^2\left(1-{r_-\over r}\right)^{2a^2\over
1+a^2}(d\theta^2+\sin^2\theta d\phi^2).
\ee
The electromagnetic vector potential 1-form is
\be
A = \sqrt{-r_+r_-\over 1+a^2}{dt\over r}.
\ee
Note that $r_-\le 0<r_+$ and so $A$ is real. We euclideanize by
setting $t=iz$ and change the sign of $r_-$ to keep the 1-form $A$
real. Since $r_-$ is now positive, we may obtain the extreme solution
by setting $r_-=r_+$ giving
$$
ds^2 = \left(1-{r_+\over r}\right)^{2\over 1+a^2}dz^2 +
\left(1-{r_+\over r}\right)^{-{2\over 1+a^2}}dr^2
$$
\be
+ r^2\left(1-{r_+\over r}\right)^{2a^2\over
1+a^2}(d\theta^2+\sin^2\theta d\phi^2),
\ee
$$
A = {r_+\over\sqrt{1+a^2}}{dz\over r}.
$$
Note that this is a Euclidean solution of the E$\overline{\rm M}$D
equations. It has a curvature singularity at $r=r_+$ and so this point
must be excluded from the space. This means that there is no need to
compactify any of the coordinates and so the topology may be taken as
${\Bbb R}^4-\{0\}$. It is again convenient to introduce isotropic
coordinates, $r=\rho+r_+$ giving
\be
ds^2 = \left(1+{r_+\over\rho}\right)^{-{2\over 1+a^2}}dz^2 +
\left(1+{r_+\over\rho}\right)^{2\over 1+a^2}(dx^2+dy^2+d\tau^2),
\label{Instant2}
\ee
where $\rho^2=x^2+y^2+\tau^2$. This then is the instanton describing
the decay process of the E$\overline{\rm M}$D vacuum. The Lorentzian
solution into which the vacuum decays is obtained by setting
$\tau=it$. The curvature singularity at $\rho=0$ now gives a singular
lightcone $x^2+y^2=t^2$ expanding radially outwards at the speed of
light from the $z$-axis. This may be viewed as the time evolution of
the initial data given by the 3-metric (setting $\tau=0$ in
(\ref{Instant2}))
\be
ds^2 = \left(1+{r_+\over\sqrt{x^2+y^2}}\right)^{-{2\over 1+a^2}}dz^2 +
\left(1+{r_+\over\sqrt{x^2+y^2}}\right)^{2\over 1+a^2}(dx^2+dy^2)
\ee
which is completely non-singular. The resulting spacetime, however, is
singular on the lightcone $x^2+y^2=t^2$~:
$$
ds^2 = \left(1+{r_+\over\sqrt{x^2+y^2-t^2}}\right)^{-{2\over
1+a^2}}dz^2
$$
\be
+ \left(1+{r_+\over\sqrt{x^2+y^2-t^2}}\right)^{2\over
1+a^2}(dx^2+dy^2-dt^2).
\ee
Thus a naked singularity spontaneously forms along the $z$-axis at
$t=0$ which then expands radially outwards at the speed of light. We
saw in section~4.3 that the action for the Euclidean instanton
(\ref{Instant2}) is zero and so one expects copious production of such
line singularities and thus that the E$\overline{\rm M}$D vacuum will
be extremely unstable. Note that this argument remains true equally
well for {\em all} values of the dilaton coupling $a$.

\sect{Conclusions}

We have derived families of black holes in theories of gravity coupled
to a Maxwell field and a scalar dilaton, in which the kinetic energies
for either or both of the fields are allowed to be negative. In the
case where just the Maxwell field is given negative kinetic energy, we
found that regular black hole solutions with zero or negative mass
were possible provided that the dilaton coupling $a$ was less than
1. If the dilaton is also given negative kinetic energy, then such
black holes are possible for {\em all} values of $a$. In this case we
also found a neutral massless wormhole solution characterized by a
non-zero scalar charge. This wormhole was described as `transparent'
because it has no horizons and timelike or null geodesics can pass
through it freely (it was shown not to exert any gravitational
attraction) and so it would be possible to see through it from one
universe into another isometric universe.

The analogue of the C-metric in these theories was derived, describing
a pair black holes accelerating away from one another. An
investigation into whether or not it is possible to satisfy the `no
strut' condition and thus remove the conical singularities of the
C-metric showed that it was possible to do so in exactly those cases
where the theory admitted zero rest mass black holes. We thus found
regular solutions describing pairs of zero rest mass black holes
accelerating away from one another. We failed, however, to find, from
these solutions, an instanton which would describe the decay of the
vacuum. However, by looking at tachyonic solutions, we were able to
give a alternative way in which the vacuum would decay. The action for
the instanton describing this instability was found to be zero so that
this decay mode is not suppressed at all.

Dyson's original argument against the convergence of the perturbation
series in QED relied on the production of pairs of oppositely charged
particles which (if the electromagnetic coupling constant were taken to
be pure imaginary) would repel one another thus destabilizing the
vacuum. In this paper, we tried to repeat the argument with pairs of
charged black holes in a theory of gravity coupled to a Maxwell field
with negative kinetic energy. We were, however, unable to find an
instanton describing the production of such pairs of black
holes. However, by using other semi-classical arguments, we were able
to show that in such a theory the vacuum is still unstable. Thus we
may still argue that a perturbative theory describing gravity and
electromagnetism cannot be uniformly convergent as a perturbative
expansion in the electromagnetic coupling constant $e$.

The results were also extended to include a scalar dilaton field and
the vacuum in this case was also found to be unstable. This result
holds for {\em all} values of the dilaton coupling $a$, in particular
for $a=-\sqrt{3}$ which is the case which arises from Kaluza-Klein
theory in which the extra dimension is taken to be timelike rather
than spacelike. This may have some relevance to recent ideas in string
theory, F-theory and other theories which consider the possibility of
extra timelike dimensions \cite{Ben}. The results imply that such
theories appear to be very unstable and, as we have seen, their solutions
may be rather pathological, as we might have expected.

\end{document}